\documentclass[twocolumn,twoside,slac_two]{revtex4}
\usepackage{graphicx}
\usepackage{fancyhdr}
\renewcommand{\headrulewidth}{0pt}
\renewcommand{\footrulewidth}{0pt}

\begin{document}

\title{High Energy Radiation from Black Holes: A Summary}

%

\author{C. D. Dermer}
\affiliation{Space Science Division, Code 7653, Naval Research Laboratory, Washington, DC 20375-5352 }
\author{G. Menon}
\affiliation{Department of Mathematics and Physics, Troy University, Troy, AL 36082}

\begin{abstract}
Bright $\gamma$-ray flares observed from sources far beyond our Galaxy 
are best explained if enormous amounts of energy 
are liberated by black holes. The highest-energy particles in nature---the ultra-high 
energy cosmic rays---cannot be confined by the Milky Way's magnetic field, 
and must originate from sources outside our Galaxy. Here we summarize 
the themes of our book, ``High Energy Radiation from Black Holes: Gamma Rays,
Cosmic Rays, and Neutrinos,"
just published by Princeton University Press. 
In this book, we develop a mathematical framework that can be used to help
 establish the nature of $\gamma$-ray 
sources, to evaluate evidence for cosmic-ray acceleration 
in blazars, GRBs and microquasars, to decide 
whether black holes accelerate the ultra-high energy cosmic rays, and 
to determine whether the Blandford-Znajek mechanism for energy extraction 
from rotating black holes can explain the differences between $\gamma$-ray blazars 
and radio-quiet AGNs. 
\end{abstract}

\maketitle

\thispagestyle{fancy}


\section{SCIENTIFIC HYPOTHESES}

In our book \cite{dm09}, a mathematical formulation of high-energy
radiation processes and strong-field gravity is presented.
This framework provides a starting point to investigate the following hypotheses:

\begin{enumerate}

\item The most energetic and powerful radiations are made by processes taking place in 
black-hole jets.

\item Ultra-high energy cosmic rays (UHECRs) are accelerated by radio and $\gamma$-ray loud blazars and GRBs.

\item Turbulence and shocks accelerate particles to high energies through Fermi processes.

\item The energy source of black holes with relativistic jets is black-hole rotation. 

\end{enumerate}

\section{ESTABLISHING THE HYPOTHESES}

We do not attempt to present a comprehensive summary of the 
observational background to be tested in our book, which is impossible 
given the rapidly expanding empirical data base arising from multiple 
facilities, in particular, the Fermi Gamma ray Space Telescope, 
the MAGIC, HESS, and VERITAS ground-based $\gamma$-ray telescopes, the Pierre Auger Cosmic Ray Observatory,
and the IceCube Neutrino Observatory. 
Instead we present a mathematical formalism to help researchers 
establish or refute the hypotheses listed above. Here we 
mention some observations that drive
the study, and related subjects dealt with in the book. 

\subsection{Black-Hole Jets as Sources of Energetic Radiations}

The Fermi telescope, building on the success of the Energetic 
Gamma Ray Experiment Telescope (EGRET) on the Compton Gamma Ray Observatory, 
found that $\sim 100$ MeV -- GeV $\gamma$ rays are produced by the blazar AGN class.
More than 30 blazar AGNs, principally of the X-ray selected BL Lac class, have now
been detected with ground-based $\gamma$-ray telescopes. 
At least 12 GRBs of both the long soft and short hard class
 have been detected with the Large Area Telescope (LAT)
on Fermi. Before that, five spark chamber events and several 
TASC (Total Absorption Shower Counter)/BATSE 
(Burst and Transient Source Experiment) events on CGRO, showed that  
GRBs produce extremely luminous multi-MeV/GeV emissions. Galactic microquasars like 
LS 5039 and Cyg X-1 are also found to be $\gamma$-ray sources.
Powerful extragalactic $\gamma$-ray sources are thought to be formed by 
black-hole jets, with the jets nearly aligned to the observer. 

To establish whether black-hole jets are the sources of luminous $\gamma$ radiations,
we develop the theory of relativistic flows, starting with relativistic kinematics
and special relativity. Compton and synchrotron processes are treated with the goal 
of presenting a framework from which to model the multiwavelength spectral energy distributions of high-energy 
radiation sources. Starting from the Compton cross section, relations 
are derived to model external 
Compton scattering involving surrounding isotropic radiation fields and anisotropic
accretion disk radiation fields. The formalism applies to scattering throughout the 
Thomson and Klein-Nishina regime, though $\delta$-function approximations are provided
to make simple back-of-the-envelope estimates.

Synchrotron radiation formulae are presented, and the synchrotron
self-Compton and synchrotron self-absorption processes are developed. The
$\gamma\gamma$ pair production process, starting from the elementary $\gamma\gamma$ cross 
section, is applied to $\gamma$-ray attenuation by target photons found in sources of 
high-energy radiation, and by target photons of the extragalactic background light.

With this formalism in hand, the nature of high-energy radiation 
sources can be examined. This includes tests for relativistic beaming 
such as the Compton catastrophe, whereby
the size scale of the source determined directly from radio observations or indirectly from the 
variability timescale of the radiation implies the level of the Compton-scattered
X-ray and $\gamma$-ray flux if the radio-through-optical/UV emission is nonthermal 
synchrotron radiation. Minimum bulk Lorentz factors
of relativistic outflows from $\gamma\gamma$
opacity arguments are derived. Moreover, minimum power requirements for
synchrotron sources can be determined from equipartition arguments. Together, these arguments
point toward relativistic outflows from highly compact sources, which almost certainly 
implicate black holes as the engines of the luminous radiation.

\subsection{UHECRs from Blazars and GRBs}

Not even the highest energy cosmic rays point back to their sources due to deflections
by the Galactic and intergalactic magnetic fields, which has made the birth 
of charged-particle astronomy difficult. In the meantime, the sources
of the UHECRs can be established indirectly 
by identifying $\gamma$-ray signatures of hadronic acceleration, and 
 directly by detecting neutrinos from their sources with, for example, 
the IceCube neutrino telescope.   
Discriminating leptonic from hadronic emission signatures in the 
absence of neutrino detection represents an important challenge
in the Fermi era. Identification of features peculiar to hadronic processes, such 
as orphan flares in blazars, represents one approach, and the 
interpretation of unusual temporal and spectral behaviors
as indicated in delayed onset or extended radiation from GRBs, represents another.

A treatment of photohadronic processes is given. This includes photomeson production 
from proton-photon interactions, ion-photon photopair (Bethe-Heitler) processes,
and ion photodisintegration. To illustrate
the use of the photohadronic physics presented in the book, a calculation is made of the UHECR spectrum from the 
superposition of long-duration GRB sources whose rate density is assumed to follow various star-formation rate functions.
(This calculation is far more difficult if UHECRs are accelerated
by blazars, insofar as supermassive black holes increase in mass with
time, their fueling is episodic, and they come from various classes of 
objects, for example, BL Lac objects and flat spectrum radio quasars.)
The associated GZK neutrino spectrum is calculated for an assumed long-duration
GRB origin of the UHECRs. The calculations are specific to UHECR protons, but 
photodisintegration cross sections and approximations are also presented, 
including giant dipole resonance and nonresonant channels. 
Calculations of neutrino production from nuclear photodisintegration is briefly described.

For completeness, binary collision processes useful for calculating radiation signatures 
from cosmic-ray interactions are given, including the processes of secondary nuclear production, bremsstrahlung, 
electron-positron annihilation, and Coulomb thermalization and knock-on electron processes.

\subsection{Particle Acceleration in Black-Hole Jets}

One of the important questions in high-energy astrophysics
is the method by which cosmic rays are accelerated to $\approx$ PeV energies
by Galactic sources, and to ultra-high energies in extragalactic sources. 
The empirical data base that must be explained
is extensive. This includes cosmic-ray anisotropy, ionic composition, 
and UHECR arrival directions. Features in the cosmic-ray spectrum include the knee at 
$\approx 3$ PeV, the second knee at $\approx 4\times 10^{17}$ eV, the 
ankle at $\approx 5\times 10^{18}$ eV, and the cutoff at $\approx 6\times 
10^{19}$ eV, probably due to GZK effects. 

A complete understanding of cosmic-ray origin requires that 
a mechanism for particle acceleration be identified. For this purpose,
an introduction to Fermi acceleration is given, starting from the two 
distinct types of Fermi acceleration mechanisms due to shock acceleration resulting
from particle diffusion across a shock front and convection downstream
of the shock, and stochastic 
acceleration by plasma wave turbulence. Dimensional arguments
leading to the Kraichnan and Kolmogorov wave turbulence spectra are given.
The Hillas criterion, whereby allowed sites for particle
acceleration to energy $E$ are restricted to those which satisfy the condition
that the Larmor radius is smaller than the size scale of the system, is presented.

First-order and second-order Fermi acceleration mechanisms are developed
to the point where maximum particle energies in different astrophysical 
environments can be derived. These maximum particle energies then 
implicate different classes of sources as plausible sites for cosmic-ray 
acceleration. Acceleration of cosmic rays by supernova remnant shocks 
remains the favored explanation for the sources which 
accelerate galactic cosmic rays, and may very well be demonstrated 
through Fermi measurements of the $\gamma$-ray spectra of supernova 
remnants. UHECR acceleration in the shocks and turbulent plasma 
formed in the relativistic jetted outflows of black-hole jets is the most 
likely explanation for their origin, compared to, for example, highly 
magnetized neutron stars or accretion or merger shocks in clusters of 
galaxies.  

The blast-wave physics developed over the past two decades to explain the 
extraordinarily luminous radiation from GRBs is also presented, including
spectral and temporal SEDs from external ahocks and colliding shells, photospheric
emission, and neutron decoupling in the expanding fireball. 

\subsection{Energy Extraction through Blandford-Znajek Processes}

If jetted black holes make the most energetic radiations in nature, with particle 
acceleration occurring via Fermi processes, a remaining question 
of fundamental importance is the energy source for these radiations. 
Accretion onto supermassive black holes obviously powers the luminous
quasi-thermal emissions in active galaxies, but less obvious are the 
reasons for the differences between radio-quiet and radio-loud 
AGNs, or for the different classes of GRBs, including X-ray flashes, 
low-luminosity GRBs, and long duration GRBs.

We consider whether these differences may be due to 
the spin of the black hole,  with the energy extracted through 
the Blandford-Znajek process. A treatment of general relativity and 
black-hole electrodynamics is given based on a ``3+1" formalism 
that provides a more compact approach than found in the original treatment. 
On this basis, we have recast the constraint equation for a force-free
black-hole magnetosphere in a form that allows one to search for
analytic solutions. We derive an exact solution which does
not give energy extraction, and an approximation solution which 
generalizes the Blandford-Znajek split monopole solution to arbitrary
values of the black-hole spin parameter. This solution is considered
in light of evidence for jetted outflows from black-hole sources.

\section{Summary}

The science covered in our book and, in particular, the 
hypotheses outlined in the first section, 
will be addressed by the 
ongoing observational activity in high-energy astronomy. 
The hypothesis that the highest energy and most luminous radiations
in nature, including the bright $\gamma$-ray flares from 
blazars and GRBs, and the UHECRs from unknown
extragalactic sources, are energized by the rotational 
energy of spinning black holes, can be tested with observatories
now in operation. This assertion will be established
or refuted as new data is collected and analyzed.

\bigskip 
\begin{acknowledgments}
This work has been supported by the Office of Naval Research and NASA.
\end{acknowledgments}

\bigskip 

\end{document}